\documentclass[12pt]{article}

\usepackage[english]{babel}
\usepackage{graphicx,rotating,epsfig}
\usepackage{a4p}

\newcommand{\TeV}{\ensuremath{\mathrm{Te\kern -0.1em V}}}
\newcommand{\GeV}{\ensuremath{\mathrm{Ge\kern -0.1em V}}}
\newcommand{\MeV}{\ensuremath{\mathrm{Me\kern -0.1em V}}}
\def\GeVc2{\ensuremath{\mathrm{ Ge\kern -0.1em V }\kern -0.2em /c^2 }}

\newcommand{\MW}{\ensuremath{M_{\mathrm{ W }}}}
\newcommand{\GW}{\ensuremath{\Gamma_{\mathrm{ W }}}}

\newcommand{\RunZ}{\hbox{Run-0}}
\newcommand{\RunI}{\hbox{Run-I}}
\newcommand{\RunII}{\hbox{Run-II}}

\begin{document}
\begin{center}
{\LARGE FERMI NATIONAL ACCELERATOR LABORATORY}
\end{center}

\begin{flushright}
%       TEVEWWG/WZ 2008/01 \\
      FERMILAB-TM-2415 \\  
% http://lss.fnal.gov/cgi-bin/getnumber.pl
%    CDF Note 9303 \\
%    D\O\ Note 5631 \\[1mm]
       1$^{\rm st}$ August 2008 \\
\end{flushright}

\vskip 1cm

\begin{center}
{\LARGE \bf Combination of CDF and D\O\ Results \\
                  on the W Boson Mass and Width\\}

\vfill

{\Large
The Tevatron Electroweak Working Group\footnote{
The Tevatron Electroweak Working group can be contacted at tev-ewwg@fnal.gov.\\
 \hspace*{0.20in} More information is available at {\tt http://tevewwg.fnal.gov}.}\\
for the CDF and D\O\ Collaborations}

\vfill

{\bf Abstract}

\end{center}

{
The results on the direct measurements of the W-boson mass and width,
based on the data collected by the Tevatron experiments CDF and D\O\
at Fermilab are summarised and combined. The CDF {\RunZ} (1988-1889) and {\RunI}
(1992-1995) results have been re-averaged using the BLUE method and
combined with {\RunI} D\O\ results and the latest published results
from CDF taken from the first period of {\RunII} (2001-2004). The
results are corrected to have consistency between the parton
distribution functions and electroweak parameters.  The resulting
Tevatron averages for the mass and total decay width of the W boson
are: $\MW = 80432\pm 39~\MeV$ and $\GW=2056 \pm 62~\MeV$.
The inclusion of a preliminary \RunII\ measurement of \GW\ from D\O\ gives
$\GW=2050 \pm 58~\MeV$.
}

\vfill

%\end{titlepage}

%\setcounter{page}{2}

\section{Introduction}

The experiments CDF and D\O, taking data at the Tevatron
proton-antiproton collider located at the Fermi National Accelerator
Laboratory, have made several direct measurements of the mass $\MW$
and total decay width $\GW$ of the W boson.  These measurements use
$e\nu$ and $\mu\nu$ decay modes.

Mass measurements have been published by CDF in
{\RunZ}~\cite{MW-CDF-RunZ},{\RunI}~\cite{MW-CDF-RUN1A, MW-CDF-RUN1B}
and recently {\RunII}~\cite{MW-CDF-RUN2} and by
D\O\ in {\RunI}~\cite{MW-D0-I,MW-D0-I-rap,MW-D0-I-edge}. Total decay
width measurements have been published by CDF in
{\RunI}~\cite{GW-CDF-Ia,GW-CDF-Ib} and recently
{\RunII}~\cite{GW-CDF-II} and by D\O~\cite{GW-D0-I} in {\RunI}. In
2004 D\O\ presented a preliminary width result based on 177
pb$^{-1}$ of \RunII\ data~\cite{GW-D0-II-0408}.

This note reports on the combination of these measurements.  The
combination takes into account the statistical and
systematic uncertainties as well as the correlations between
systematic uncertainties, and replaces our previous
combinations~\cite{MWGW-RunI-PRD,GW0508}.  The measurements are
combined using a program implementing a numerical $\chi^2$
minimization as well as the analytic BLUE method~\cite{Lyons:1988,
Valassi:2003}. The two methods used are mathematically equivalent, and
are also equivalent to the method used in a previous
combination~\cite{MWGW-RunI-PRD,GW0508}.  In addition, the BLUE method yields
the decomposition of the error on the average in terms of the error
categories specified for the input measurements~\cite{Valassi:2003}.

This analysis has three significant changes over previous averages:
\begin{itemize}
\item The individual $e\nu$ and $\mu\nu$ results for CDF Run-0, Run-Ia
and Run-Ib data are now combined for each run period using the BLUE
method to achieve a consistent statistical treatment across all the
results. The \RunI\ D\O\ and  \RunII\ CDF results were already internally
combined using the BLUE method.
\item For the mass measurements, the central values are corrected to
use the same parton distribution functions (PDFs) and the same SM
value for the W width. The W mass error arising from an uncertainty in
the W width is thus now consistently treated across all measurements.
\item For the width measurements, the values are corrected back to the
same assumed W mass value to achieve consistency across all results.
\end{itemize}
The values of \MW\ and \GW\ quoted here correspond to a
definition based on a Breit-Wigner denominator with a mass-dependent
width, $|M^2-\MW^2+iM^2\GW/\MW|$.

\section{Mass of the W Boson}

The five measurements of $\MW$ to be combined are given in
Table~\ref{tab:MW-inputs}. The CDF \RunZ, Run-Ia and Run-Ib values are
each themselves averages of two individual measurements where
internal correlated systematic errors e.g. momentum scale, are
accounted for in the averaging. The \RunI\ D\O\ measurement combines
10 individual measurements using the BLUE method. The \RunII\ CDF
measurement combines six individual measurements using the BLUE method.
The early CDF measurements from \RunZ\ and \RunI\ were not combined using the BLUE method,
instead a simpler formulation was used. These
measurements were combined using only the uncorrelated errors, and
then the correlated errors were added in quadrature. In the case here,
where the correlated errors are small with positive correlation
coefficients, this method gives a very similar result to the BLUE method but we
have taken this opportunity to re-combine the CDF \RunZ\ and \RunI\
results using the BLUE method to achieve a consistent statistical
treatment across all the results.  
%%%
Firstly, the combination of $e\nu$ and $\mu\nu$ results for the {\RunZ}, Run-Ia and Run-Ib CDF data are combined internally using the BLUE method. This changes the {\RunZ}, Run-Ia and Run-Ib CDF \MW\ values by  
$-3.5$~MeV, $-3.5$~MeV and $0.1$~MeV respectively. These BLUE corrections are listed in
Table~\ref{tab:MW-inputs}. When these corrected values are further combined using the BLUE method, the Run-0/I CDF combination is changed from $80433 \pm\ 79$~MeV quoted
in~\cite{MW-CDF-RUN1B} and used in previous
combinations~\cite{MWGW-RunI-PRD} to $80436 \pm\ 81$~MeV. This combination uses the same PDF corrections (see below) and width errors as used in the original CDF Run-0/1 combination.

In fitting for \MW\ from the measured transverse mass, charged lepton and neutrino $p_{\rm T}$ distributions 
a fixed value for \GW\ is used. A variety of
values have been used in the five measurements. Here we have corrected
the mass values so that they all use the SM prediction of $\GW=2093 \pm
2$~MeV~\cite{PR} corresponding to $\MW = 80399 \pm 25$~MeV which is
the world average resulting from this latest combination.
The results are corrected using the relation $\Delta\MW = -(0.15 \pm
0.05) \times \Delta\GW$ which is the average, including the variance, of the empirical shifts in \MW\ determined by CDF and D\O\ when \GW\ is varied. The \MW\ uncertainty from the 2~MeV uncertainty~\cite{PR} in the SM value of \GW\ is 0.3~MeV (which we round to 0.5 MeV); this is added in quadrature with the uncertainty in the \GW\ correction i.e. $0.05 \times \Delta\GW$ and used across all measurements. Previous quoted \GW\ errors, where they existed, are
subtracted in quadrature from the total error. $9.9$ ~MeV is subtracted
in quadrature from the D\O\ \RunI\ combination, and $19.9$~MeV is subtracted in quadrature 
from the CDF Run-Ia result. 0.5~MeV is added in quadrature to the CDF {\RunZ}, Run-Ib and \RunII\
combinations since these three combinations did not include an
uncertainty from \GW. The \GW\ correction is listed in Table~\ref{tab:MW-inputs}.

The CDF \RunZ\ and Run-Ia results were obtained from very old PDF sets 
(MRS-B~\cite{MRS-B} and MRSD-$^\prime$~\cite{MRS-D} respectively)
that did not utilise the W charge asymmetry results and so 
provide somewhat offset predictions from the more modern PDF sets used
in the later analyses. The predictions based on the more modern
MRS~\cite{MRS} and CTEQ~\cite{CTEQ} sets used in Run-Ib and \RunII\ analyses 
have a variance smaller than the common PDF errors assumed in these analyses ie
$\approx 10$~MeV. We therefore only apply PDF corrections to the CDF
\RunZ\ and Run-Ia results since the shifts for these data are larger
than 10~MeV. Corrections of $+20$~MeV and $-25$~MeV are applied to the
\RunZ\ and Run-Ia published results respectively. We however retain the
PDF uncertainties of 60~MeV and 50~MeV quoted in the original
publications.  
We note that these corrections were also applied in the
\RunI\ CDF combination presented in~\cite{MW-CDF-RUN1B}. 
The realisation of a common PDF uncertainty, utilising modern PDFs, will be the subject of a future study.

The \MW\ values and revised errors after these three corrections (\GW, PDF and BLUE)
are listed as ``\MW\ (corrected)'' and ``Total
BLUE Error (\GW\ corrected)'' in Table~\ref{tab:MW-inputs}. After these
corrections the following experiment averages are determined:
\begin{itemize}
\item CDF {\RunI}, $\MW = 80436 \pm 81$~MeV to
be compared with $\MW = 80433 \pm 79$~MeV in~\cite{MW-CDF-RUN1B}. 
\item CDF {\RunI}/II, $\MW = 80421 \pm 43$~MeV 
to be compared with $\MW = 80418 \pm 42$~MeV quoted in \cite{MW-CDF-RUN2}. 
This shift is almost entirely due to the $3$~MeV change resulting from the consistent use of the BLUE method across the \RunZ\ and \RunI\ CDF results.
\item D\O\ \RunI\ and hence D\O\ average, $\MW = 80478 \pm 83$~MeV compared to $\MW = 80483 \pm 84$~MeV quoted in \cite{MW-D0-I-edge}. This change is due to the \GW\ correction.
\end{itemize}

Three systematic errors are assumed to be fully correlated between all
measurements, namely: (i) parton distribution functions and parton
luminosity (PDF), (ii) Electroweak radiative corrections (EWK RC), 
and (iii) the width of the W boson ($\GW$). Further
details on the sources of systematic uncertainties are given in~\cite{MWGW-RunI-PRD} 
and the individual publications of the
two experiments~\cite{MW-CDF-RunZ,MW-CDF-RUN1A,
MW-CDF-RUN1B,MW-CDF-RUN2,MW-D0-I,MW-D0-I-rap,MW-D0-I-edge}.

\begin{table}[htbp]
\begin{center}
\renewcommand{\arraystretch}{1.30}
\begin{tabular}{|l||r|r|r|r||r|}
\hline       
       &  \multicolumn{1}{|c|}{{Run-0}} 
       &  \multicolumn{3}{|c||}{{\RunI}} 
       &  \multicolumn{1}{|c|}{{\RunII}} \\
\hline
       & \multicolumn{1}{|c|}{ CDF } 
       & \multicolumn{1}{|c|}{ CDF-Ia }
       & \multicolumn{1}{|c|}{ CDF-Ib }
       & \multicolumn{1}{|c||}{ D\O\ } 
       & \multicolumn{1}{|c|}{ CDF } \\
\hline       
\hline 			       
\MW\ published                                        &  79910      & 80410     & 80470      & 80483        & 80413 \\
Total Error published                                 &  390         & 180         & 89            & 84              & 47.9  \\
\GW\ used in publication                            &  2100        & 2064       & 2096        &  2062         & 2094 \\ 
\GW\ correction applied                              &   1.1          &  $-$4.4    &  0.5         & $-$4.7        & 0.2 \\
PDF correction applied                               &   20        & $-$25      & 0             & 0                 & 0 \\ 
BLUE correction applied                             & $-$3.5       & $-$3.5     & $-$0.1     & 0                & 0 \\ \hline
\MW\ (corrected)                                         & 79927.6    & 80377.1  & 80470.4   & 80478.3     & 80413.2 \\   
Total BLUE error (\GW\ corrected)              & 390.9        & 181.0     & 89.3          & 83.4           & 47.9 \\
Uncorrelated BLUE error (\GW\ corrected) & 386.1        &	172.8   & 87.9           &  82.1	     & 44.7 \\
\hline                         
\hline                         
PDF Error  (published/this analysis)           &    60          &    50        &     15         &     8.1       &    12.6  \\
EWK RC  Error  (published/this analysis)   &    10          &    20        &     5           &    12         &    11.6 \\
 $\GW$ Error (published)                            &     0           &    20        &     0           &     10        &     0 \\
 $\GW$ Error (this analysis)                        &    0.5         &   1.5        &     0.5            &     1.5    &    0.5 \\ \hline \hline
\end{tabular}
\end{center}
\caption[Input measurements]{Summary of the five measurements of $\MW$
performed by CDF and D\O. All numbers are in $\MeV$.  The published
values and the corrected values used in the average are shown. The
three sources of correlated systematic error (PDF, EWK RC, \GW) are
explicitly given.}
\label{tab:MW-inputs}
\end{table}

The combined Tevatron value for the W-boson mass is:
\begin{eqnarray}
\MW & = & 80432\pm39~\MeV\,,
\end{eqnarray}
where the total error of $39~\MeV$ contains the following components:
an uncorrelated error of $35~\MeV$; and correlated systematic error
contributions of: parton distribution functions $13~\MeV$, electroweak
radiative corrections $11~\MeV$ and W-boson width $0.7~\MeV$, for a total
correlated systematic error of $17~\MeV$. 
The global correlation matrix for the 5 measurements is shown in Table~\ref{tab:MW-corr}.
\begin{table}[htbp]
\begin{center}
\renewcommand{\arraystretch}{1.30}
\begin{tabular}{|l||c|c|c|c||c|}
\hline       
       &  \multicolumn{1}{|c|}{{Run-0}} 
       &  \multicolumn{3}{|c||}{{\RunI}} 
       &  \multicolumn{1}{|c|}{{\RunII}} \\
\hline
       & \multicolumn{1}{|c|}{ CDF } 
       & \multicolumn{1}{|c|}{ CDF-Ia }
       & \multicolumn{1}{|c|}{ CDF-Ib }  
       & \multicolumn{1}{|c||}{ D\O\ } 
       & \multicolumn{1}{|c|}{ CDF } \\
\hline       
\hline 			       
CDF Run-0      &  1.0      &  0.05     &     0.03    &   0.02      &  0.05 \\
CDF-Ia             &            &   1.0      &      0.05    &   0.04      &  0.10  \\
CDF-Ib             &            &             &     1.0        &   0.02      &  0.06  \\
D\O\                  &            &            &                  &   1.0        &  0.06  \\
CDF-II               &            &           &                   &                &  1.0  \\ 
\hline
\end{tabular}
\end{center}
\caption[Input measurements]{Matrix of global correlation coefficients between the 5 measurements
of Table~\ref{tab:MW-inputs}.}
\label{tab:MW-corr}
\end{table}

\begin{figure}[htbp]
\begin{center}
\includegraphics[width=0.8\textwidth]{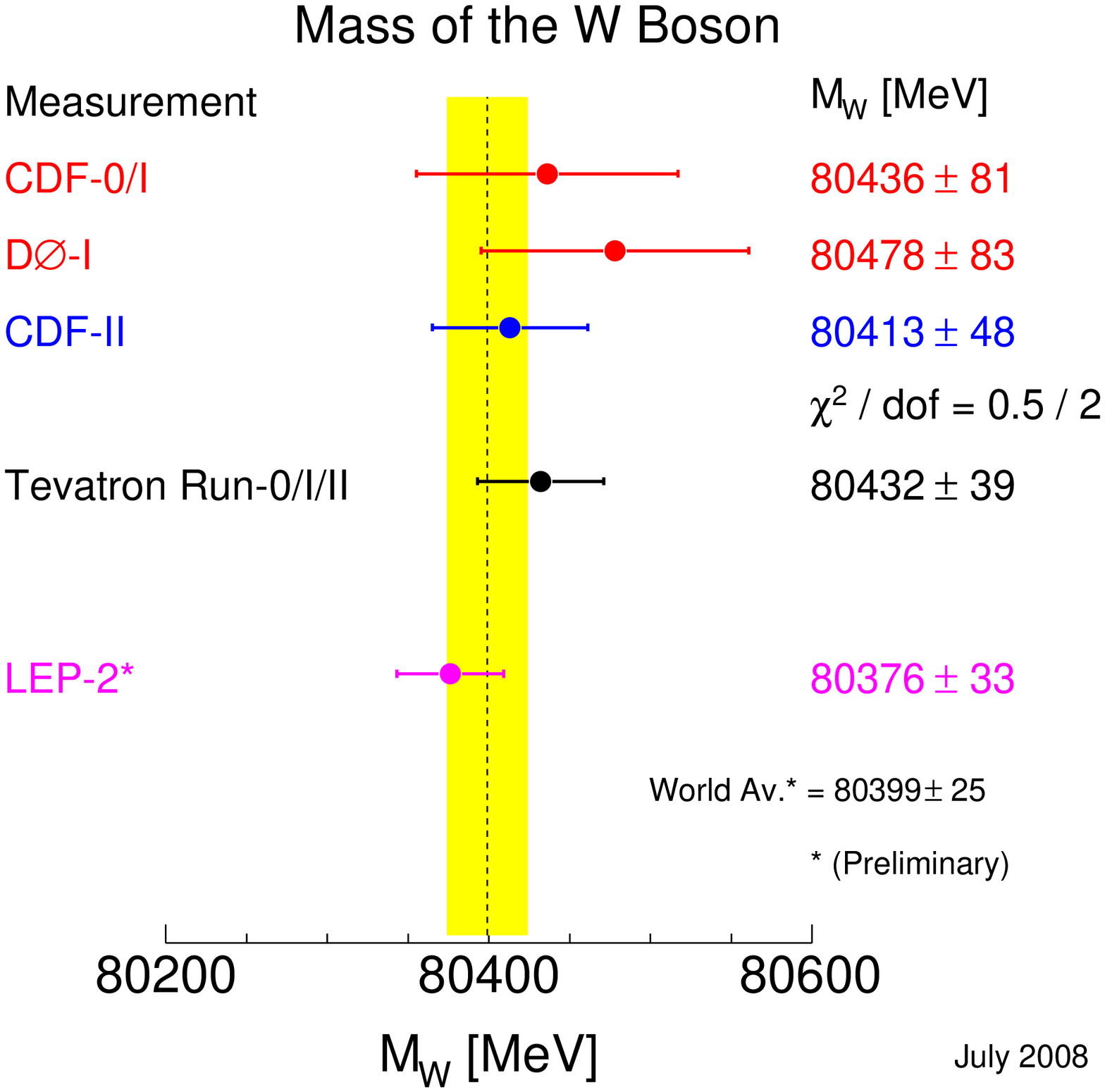}
\end{center}
\caption[Comparison of the measurements of the W-boson mass]
{Comparison of the measurements of the W-boson mass and their average.
The most recent preliminary result from LEP-2~\cite{LEPEWWG07} is
also shown. The Tevatron values shown are the corrected values.}
\label{fig:mw-bar-chart} 
\end{figure}

The $\chi^2$ of this average is 2.4 for 4 degrees of freedom,
corresponding to a probability of 66\%, showing that all measurements
are in good agreement with each other which can also be seen in
Figure~\ref{fig:mw-bar-chart}.  The $\chi^2$ with respect to the Tevatron average 
of the three experiment averages (CDF-I/O, D\O-I, CDF-II) shown in 
Figure~\ref{fig:mw-bar-chart} is 0.5 for 2 degrees of freedom. 

A combination with the latest LEP-2 value~\cite{LEPEWWG07} yields 
a world average of \MW = 80399 $\pm$ 25~MeV.

\section{Width of the W Boson}

\begin{table}[htbp]
\begin{center}
\renewcommand{\arraystretch}{1.30}
\begin{tabular}{| l | r || r | r || r | r | }
\hline       
       &  \multicolumn{3}{|c||}{{\RunI}} 
       &  \multicolumn{2}{|c|}{{\RunII}} \\
\hline
       & \multicolumn{1}{|c|}{ CDF-Ia }
       & \multicolumn{1}{|c|}{ CDF-Ib }
       & \multicolumn{1}{|c||}{ D\O-Ib}
       & \multicolumn{1}{|c|}{ CDF } 
       & \multicolumn{1}{|c|}{ D\O\ }  \\
\hline       
\hline 			       
\GW\ published                                           & 2110      & 2042.5      &  2231    & 2032     & 2011 \\
Total Error published                                   &  329      &  138.3       &  172.8     & 72.4     & 142.1  \\
\MW\ used in publication                             & 80140    & 80400    & 80436  &  80403   & 80426 \\ 
\MW\ correction applied                              & $-$78    & 0.3         &  11.1    &   1.2       &   8.1 \\
%BLUE correction applied                          & 0           &  0           &  0         &    0         &    0   \\
%PDF   correction applied                           & 0            &  0          &  0         &    0          &   0   \\  
\GW\ (corrected)                                          & 2032     & 2042.8  & 2242.1  & 2033.2    &  2019.1 \\ \hline   
Total BLUE error (corrected)                        & 329.3    & 138.3    & 172.4   &    72.4       & 141.6 \\ 
Uncorrelated BLUE error (\GW\ corrected) & 327.6        &	136.8   & 167.4  &  68.7	    & 138.7 \\ \hline \hline
PDF Error (published)                       &    0      &    15      &    39     &     20       &    27  \\
PDF Error (this analysis)                   &    15      &    15      &    39     &     20      &    27  \\
EWK RC  (published/this analysis)    &    28      &    10     &     10    &    6          &    3 \\ 
\MW\ Error (published)                      &    0      &    10      &    15     &     9     &    15  \\
\MW\ Error (this analysis)                  &    9       &     9      &      9      &      9        &   9  \\ \hline \hline
\end{tabular}
\end{center}
\caption[Input measurements]{Summary of the five measurements of
$\GW$ performed by CDF and D\O. All numbers are in $\MeV$.  The
published values and the corrected values used in the average are
shown. The three sources of correlated systematic error (PDF, EWK RC,
\MW) are explicitly given.}
\label{tab:GW-inputs}
\end{table}

As for the mass combination, we have made two corrections to achieve
consistency across all the width results. The CDF Run-Ib results have been
recombined using the BLUE method and all results have been corrected
so that the values correspond to the same assumed $\MW = 80399 \pm
25$~MeV. 
The use of the BLUE method results in a negligible change to the CDF Run-Ib result.
No PDF corrections have been applied. 
The results are corrected to a consistent mass value of \MW\ = 80399~MeV using the relation 
$\Delta\GW = -(0.3 \pm 0.1) \times \Delta\MW$. 
This is the CDF and D\O\ average of the empirically determined shift in \GW\ when \MW\ is varied.
A common error of 9~MeV from the uncertainty in \MW\ is now used across all
measurements and supercedes the published \GW\ error. This 9~MeV error encompasses the 25~MeV uncertainty in \MW\ and the 0.1 uncertainty in d\GW/d\MW. 4.35~MeV is subtracted from the CDF-Ib results in quadrature and 12~MeV in quadrature from the D\O\ results. After these corrections, we obtain the following experiment averages:
\begin{itemize}
\item CDF {\RunI}, $\GW = 2041 \pm 128$~MeV
\item CDF Run-I/II, $\GW = 2035 \pm 64$~MeV
\item Preliminary D\O\ Run-I/II,  $\GW = 2108 \pm 112$~MeV
\end{itemize}

\begin{figure}[bp]
\begin{center}
\includegraphics[width=0.8\textwidth]{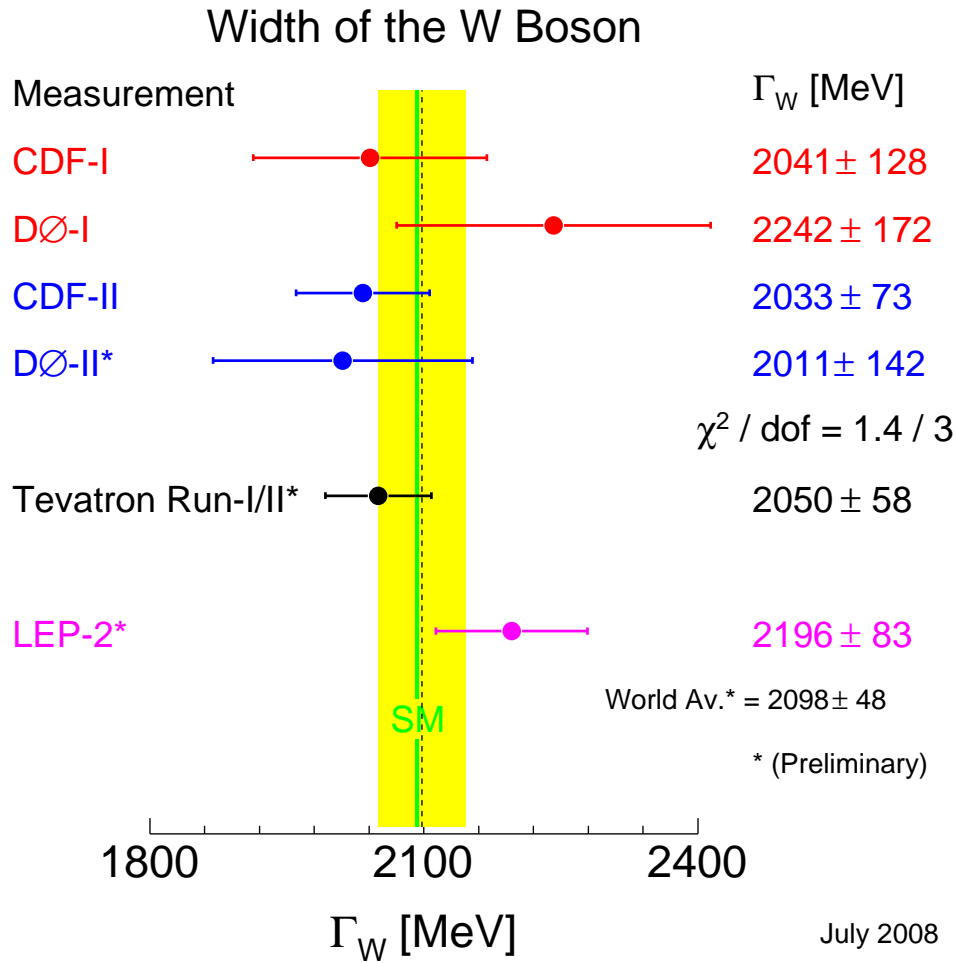}
\end{center}
\caption[Comparison of the measurements of the W-boson width]
{Comparison of the measurements of the W-boson width and their
average.  The most recent preliminary result from
LEP-2~\cite{LEPEWWG07} and the Standard Model prediction are also
shown. The Tevatron values shown are the corrected values.}
\label{fig:gw-bar-chart} 
\end{figure}

The combined Tevatron value for \GW\ is:
\begin{eqnarray}
\GW & = & 2056\pm62~\MeV\,,
\end{eqnarray}
from published results and $\GW = 2050 \pm 58$~MeV if the preliminary D\O\ result is included. 
The combination of the published Tevatron results has a $\chi^2$ of 1.3 (1.4) for 3 (4)
degrees of freedom, corresponding to a probability of 72\% (84\%) depending of whether the preliminary 
D\O\ result is included or not. All measurements are in good agreement with each other and this can
also be seen in Figure~\ref{fig:gw-bar-chart} where the CDF-I, CDF-II, D\O-I and D\O-II averages are shown which has a $\chi^2$ of 1.4 for 3 degrees of freedom with respect to the Tevatron average 
including the D\O\ preliminary result. 

The combination of published results 
which has a total error of 62~MeV contains the following components: an uncorrelated error of 
57 MeV; and correlated systematic error contributions of: parton distribution functions 21 MeV, 
electroweak radiative corrections 8 MeV, and W-boson mass 9 MeV, for a total correlated systematic 
error of 24 MeV. 
The global correlation matrix for the 5 measurements is shown in Table~\ref{tab:GW-corr}.
Figure~\ref{fig:gw-bar-chart}

\begin{table}[htbp]
\begin{center}
\renewcommand{\arraystretch}{1.30}
\begin{tabular}{|l||c|c|c||c|c|}
\hline       
       &  \multicolumn{3}{|c||}{{\RunI}} 
       &  \multicolumn{2}{|c|}{{\RunII}} \\
\hline
       & \multicolumn{1}{|c|}{ CDF-Ia }
       & \multicolumn{1}{|c|}{ CDF-Ib }
       & \multicolumn{1}{|c||}{ D\O-Ib}
       & \multicolumn{1}{|c|}{ CDF } 
       & \multicolumn{1}{|c|}{ D\O\ }  \\
\hline       
\hline 			       
CDF-Ia     &  1.0      &  0.01     &     0.02      &   0.02      &  0.01 \\
CDF-Ib     &           &   1.0     &     0.03      &   0.04      &  0.03  \\
D\O-I      &           &           &     1.0       &   0.07      &  0.05  \\
CDF-II     &           &           &               &   1.0       &  0.06  \\
D\O-II    &            &           &               &             &  1.0  \\ 
\hline
\end{tabular}
\end{center}
\caption[Input measurements]{Matrix of global correlation coefficients between the 5 measurements
of Table~\ref{tab:GW-inputs}.}
\label{tab:GW-corr}
\end{table}

 A combination with the latest LEP-2 value~\cite{LEPEWWG07} gives a preliminary 
world average of $\GW = 2098 \pm 48$~MeV which is in excellent agreement with the SM 
prediction of $\GW = 2093 \pm 2$~MeV~\cite{PR}.

\section{Summary}

Combinations of the direct CDF and D\O\ measurements of the mass and
total decay width of the W boson are presented. Corrections have
been made to achieve a consistent treatment across all
measurements and the values are corrected to consistent values for the
SM parameters. The Tevatron averages of published results are: $\MW =
80432\pm39~\MeV$ and $\GW=2056\pm62~\MeV$ or $\GW=2050\pm58~\MeV$ if
preliminary results are included.

\end{document}